\documentclass[article, a4paper]{amsart}

\usepackage{amsmath}
\usepackage{latexsym}
\usepackage[all]{xy}
\usepackage{color}
\newtheorem{teorema}{Theorem}[section]
\newtheorem{definicion}[teorema]{Definition}
\include{amslatex}
\newtheorem{proposicion}[teorema]{Proposition}

\newtheorem{corolario}[teorema]{Corollary}

\newtheorem{ejemplo}[teorema]{Example}

\numberwithin{equation}{section}
\begin{document}

\begin{title}[Second order equation for a point charged particle]
{ A second order differential equation for a point charged particle}
\end{title}
\maketitle
\author{

\begin{center}

Ricardo Gallego Torrom\'e\\
Departamento de Matem\'atica\\
Universidade Federal de S\~ao Carlos, Brazil\\
{Email: rigato39@gmail.com}
\end{center}}

\begin{abstract}
A model for the dynamics of a classical point charged particle interacting with higher order jet fields is introduced. In this model, the dynamics of the charged particle is described by an implicit ordinary second order differential equation. Such equation is free of run-away and pre-accelerated solutions of Dirac's type. The theory is Lorentz invariant, compatible with the first law of Newton and Larmor's power radiation formula. Few implications of the new equation in the phenomenology of non-neutral plasmas is considered.
\end{abstract}

\section{Introduction}

To find a consistent theory of classical electrodynamics of point charged particles interacting
with its own radiation field  is one of the most notable  open problems in classical field theory.  Historically, the investigation of the radiation reaction of point charged particles lead to the relativistic Abraham-Lorentz-Dirac  equation \cite{Dirac} (in short, the ALD equation). However, the  theoretical problems and paradoxes associated with ALD inclined  many authors to support the view that the classical electrodynamics of point charged particles should not be based on the coupled system ALD /Maxwell system of equations. Attempts to find a consistent theory of classical electrodynamics of point particles include reduction of order schemes for the ALD equation \cite{LandauLifshitz}, re-normalization group schemes \cite{Spohn2},
  higher order derivative field theories \cite{Bopp, Podolsky}, Feynman-Wheeler's absorber theory of electrodynamics \cite{FeynmanWheeler},
 models where the observable mass is variable with time \cite{Bonnor, Larmor}, non-linear electrodynamic theories \cite{BornInfeld} and dissipative force models \cite{MoPapas, Herrera1990a}.

In this  paper we investigate  if in the framework of classical physics it is  possible to have  a consistent description of the dynamics of fields and point charged particles that takes into account the radiation-reaction effects. This is not a trivial question and it can be the case that the problems of the ALD are inherent to the classical description of point charged particles,
since at a more fundamental level, the description of a charged particle interacting with an external field should be considered in the framework of relativistic quantum field theory. However, we do not pursue here a quantum mechanical treatment of the problem. This is because our question, by its own character, should be discussed at the classical level. Indeed, we found a positive answer to the question.

The fact that the ALD equation is derived from fundamental principles of classical field theory suggests that it is indeed the {\it correct theoretical equation} in such framework. Therefore, if one tries to obtain a consistent theory without the problems that the ALD equation has within the framework of classical physics, it is necessary to {\it relax} the fundamental assumptions of  classical field theory.

In this paper we explore a new possibility based on the mathematical framework introduced in \cite{Ricardo2012} of geometry of high order jet fields  to obtain an implicit second order differential equation for point charged particles  free of the
run away and pre-accelerated solutions of the ALD equation. The new equation of motion is based upon four assumptions. First, that the classical electromagnetic field is described by a {\it generalized higher order field}. The most direct  application of this assumption leads to a linear higher order jet electrodynamics. Second, the assumption that proper acceleration of a point charged particle is bounded, an idea that lead us to use  {\it maximal acceleration geometry} \cite{Ricardo2015}. Third, the  new theory must be consistent with a higher order generalization of general relativity, according to the view that classical fields are higher order jet fields. Fourth, the assumption that in this generalized setting Larmor's radiation  law is still approximately valid. However, this assumption is heuristically motivated in this paper. Adopting these fundamental ideas and by the use of elementary methods taken from  differential geometry, a new equation of motion is obtained that is of second order, since the {\it Schott's term} does not appear in the new equation. The elimination of the Schott's term in the equation of motion is not new. Indeed, J. Larmor and later  W. Bonnor investigated a theory that exactly does this. In their theories, the {\it observable rest mass} of the point charged particle could vary with time, being this the source for the radiation reaction force. On the other hand, the {\it bare mass} was constant. In contrast, in our proposal the bare mass can vary during the evolution, but the observable rest mass is constant. In our theory the higher order electromagnetic fields contain additional degrees of freedom associated to the higher order jet corrections.

The new differential equation \eqref{equationofmotion} is compatible with Newton's  first
law of dynamics and it does not have the un-physical solutions that the original ALD equation has. Since we impose  Larmor's radiation law to hold as well, it is reasonable to say that the new equation of motion is compatible with energy-momentum conservation in flat spacetime. The equation can be written in general covariant form (the general covariant version of the equation \eqref{equationofmotion} is \eqref{covariantequationofmotion}). Moreover, the equation \eqref{equationofmotion} can be approximated by an ordinary differential equation where the second time derivative is explicitly isolated. We argue that in the theoretical domain of applicability of the theory, the approximate equation is equivalent to the original new equation of motion. This opens the possibility to use standard ODE theory to prove existence and uniqueness, bounds and regularity properties of the solutions for the approximated equation.

However, the derivation of equation \eqref{equationofmotion} breaks down in two dynamical regimes.
The first corresponds  when the system is near to a {\it curve of  maximal acceleration}. Indeed, the domain of applicability of the theory corresponds to world lines far from the maximal acceleration regime. The second corresponds to {\it covariant uniform acceleration}. In the case of covariant uniform acceleration,  a similar analysis as in the non-uniform acceleration case leads us to a differential equation which is the Lorentz force equation with a {\it constant total electromagnetic field}. We discuss the significance of this result.

 The structure of this paper is the following. In {\it section} 2, the notion of generalized higher order electromagnetic field is introduced. We have developed a local treatment, since this is enough for our objective of finding an ODE for point charged particles. The necessary notation and notions from jet bundle theory are collected in {\it Appendix} A  in a way that the reader un-familiar with the theory can follow the details of the argument. The fundamental equations of the generalized Maxwell's theory are briefly presented. In {\it section 3}, the basic notions of the geometry of metrics of maximal acceleration are reviewed. A perturbation scheme in powers of the quotient proper acceleration/maximal acceleration and its derivatives is also introduced. We also discuss the kinematical constraints for on-shell particles when the spacetime structure is determined by a metric of maximal acceleration.  {\it Section 4} gives a short derivation of the ALD equation.  The method  is based on the use of adapted frames and imposing a consistency  condition between the new differential equation and the relativistic Larmor's radiation law. In {\it section 5}, it is discussed Larmor's relativistic radiation law in the contest of linear, higher order jet electrodynamics in a maximal acceleration spacetime. Then following a similar argument as in {\it section} 4, the new equation of motion \eqref{equationofmotion} is obtained. We discuss several properties of the new equation, including the absence of run away solutions, the natural value of the maximal acceleration for the point charged particle, the case of constant covariant acceleration and the absence of pre-accelerated solutions of Dirac's type. The value of the maximal acceleration for a point charged particle with an electromagnetic field is derived. This leads to the possibility of experimentally test the theory.  A short discussion of the theory is presented in {\it section 6}. In {\it Appendix A}, several fundamental definitions and results from jet theory and higher order geometry are presented.
\\

{\bf Notation.} In this paper $M$ is the  four dimensional spacetime manifold.
$TM$ is the tangent bundle of $M$ and $\eta$ is a fixed background Lorenztian metric defined on $M$, although by practical purposes we  will often assume that $\eta$ is the Minkowski metric with signature $(-1,1,1,1)$. The null cone bundle of $\eta$ is $\hat{\pi}:NC\to M$. Local coordinates on $M$ are denoted by
$(x^0,x^1,x^2,x^3)$ or simply by $x^{\mu}$. It will be useful  to identify notation for
points and coordinates, since we are working open domains of $M$ homeomorphic to $\mathbb{R}^4$ and we are considering local properties of the dynamical systems. Natural coordinates on $TM$ and $NC$ are denoted by $(x,y)$.
 For the notation and basic definitions of jet bundle theory we refer to {\it Appendix} A. In this work we have use natural units where the speed of light in vacuum is $c=1$.
\section{Notion of generalized higher order fields}

 For classical fields, the simplest model of test particle is the
  point charged particle, described by a $1$-dimensional  world line sub-manifold   $x:I\to M$.
   By analyzing the departure of the world-line from being a geodesic of the Lorentzian metric $\eta$ and with a given theoretical model for the interaction of the charged particle and the classical fields, one can infer information on the  fields from observation. In general, it is necessary to consider more than one test particle to completely determine the value of the classical field at a given point of the spacetime manifold. Therefore, a classical field can be thought as a
     functional from the   {\it path space of world-lines of physical test particles} in a convenient geometric target space.

In such contest jet bundle theory is the natural mathematical framework, since it deals systematically with Taylor's expansions of
 functions and sections of a given differentiable regularity. Moreover, jet bundle theory is
 useful when dealing with geometric objects  defined along maps, independently of their geometric character and it is a theory extensively used in problems of calculus of variations (see {\it Appendix A} and references therein) and more generally, in the formal theory of partial differential equations.

  Jet fields have  well defined transformation rules under local coordinate transformations and under reasonable assumptions, the smooth sections of a vector bundle $\mathcal{E}\to M$ can be approximated locally by jet sections. This is as a consequence of  Peetre's type theorems (see \cite{KolarMichorSlovak}, pg. 176). Thus for a particular class of operators  on arbitrary bundles (non-increasing operators, \cite{KolarMichorSlovak}), it is equivalent to work with operators acting on finite order jet fields than with the sections and operations of the original theory. The non-increasing assumption on the operators is in concordance with the absence of run-way solutions of the theory that we will consider, making even more natural the use of jet theory in this contest.

\subsection{Notions of generalized electromagnetic field and current}
\
We  adopt here the following generalizations of the Faraday and the excitation tensors and of the electromagnetic density current,
\begin{definicion}
The electromagnetic field $\bar{F}$ along the lift $^kx:I\to J^k_0(M)$ is a $2$-form  that in local natural coordinates can be written as
\begin{equation}
\bar{F}(\,^kx)=\bar{F}(x,\dot{x},\ddot{x},\dddot{x},..., x^{(k)})=\big(F_{\mu\nu}(x)+
\Upsilon_{\mu \nu}(x,\dot{x},\ddot{x},\dddot{x},..., x^{(k)})\big)d_4 x^{\mu}\wedge d_4 x^{\nu}.
\label{electromagneticfield}
\end{equation}

The excitation tensor $\bar{G}$ along the lift $^kx:I\to J^k_0(M)$ is a $2$-form
\begin{equation}
\bar{G}(\,^kx)=\bar{G}(x,\dot{x},\ddot{x},\dddot{x},..., x^{(k)})=\big(G_{\mu\nu}(x)+
\Xi_{\mu \nu}(x,\dot{x},\ddot{x},\dddot{x},..., x^{(k)})\big)d_4 x^{\mu}\wedge d_4 x^{\nu}.
\label{excitationtensor}
\end{equation}

The density current $\bar{J}$ is represented by a $3$-form
\begin{equation}
\bar{J}(x,\dot{x},\ddot{x},\dddot{x},..., x^{(k)})=\,\big(J_{\mu\nu\rho}(x)\,+
\Phi_{\mu\nu\rho}(x,\dot{x},\ddot{x},\dddot{x},..., x^{(k)})\big)\,d_4x^{\mu}\wedge d_4x^{\nu}\wedge d_4x^{\rho}.
\label{generalizedcurrent}
\end{equation}
\end{definicion}
 To each of these three types of generalized fields there are associated standard fields $F,G\in \,\Lambda^2 M$ and $J\in\,\Lambda^3M$ defined locally by the components $F_{\mu\nu}(x)$, $G_{\mu\nu}(x)$ and $J_{\mu\nu}(x)$ respectively.

 A generalized field associates to each pair of tangent vector fields $e_\mu,\,e_\nu$ along the curve $x:I\to M$ an element of the $k$-jet along $x$, for instance,
 \begin{align*}
 \big(e_\mu(x(\tau)),\,e_\nu(x(\tau))\big)\mapsto \bar{F}(\,^kx(\tau)).
 \end{align*}
 The fact that the fields take values on higher order jet bundles over the spacetime manifold
 is the mathematical implementation of our idea that fields depend
  on the state of motion of the test particle in a fundamental way. At this stage we do not fix the value of the order $k$.
  We will keep it free until in the following {\it sections} it will be fixed by physical arguments to be $k=2$.

There is a coordinate-free formulation for the generalized fields \cite{Ricardo2012}. In particular, the Hodge star operator $\star$ and the
nilpotent exterior derivative $d_4$ are well defined geometric objects acting on generalized forms.
With the aid of the nilpotent  {\it exterior derivative} $d_4$ and the Hodge star operator
$\star$ defined by the Lorentzian metric $\eta$, the homogeneous of Maxwell's equations is the expression
\begin{align}
d_4\bar{F}=0.
\label{homogeneousequation}
\end{align}
If we assume the constitutive relation $\bar{G}=\,\star \bar{F}$, then the generalized inhomogeneous Maxwell's equations are
\begin{align}
d_4\star \,\bar{F}=\,{J}+d_4\,\star \Upsilon.
\label{nonhomogeneous equation}
\end{align}

From equations \eqref{homogeneousequation} and \eqref{nonhomogeneous equation}
it is possible to construct an effective theory which is equivalent to the standard Maxwell's theory. In particular, the Faraday tensor $F$ must be such that
\begin{align}
d F=0
\label{equationforF}
\end{align}
holds good; $G$ must be a solution of the equation
\begin{align}
d\star\, F=\,J
\label{equationfor*F}
\end{align}
and the conservation of the current density are
\begin{align}
d{J}=0.
\label{equationforJ}
\end{align}
The boundary conditions for $F$ and $G$ are contained in the date determining the boundary conditions for $\bar{F}$ and $\bar{G}$ respectively.

\section{Maximal acceleration geometry}

\subsection{Maximal acceleration and electrodynamics}
\
The general idea of maximal acceleration in electrodynamics is not new.
Maximal acceleration in the framework of Newton theory was used by Lorentz in his theory of the electron, where it appears as a constraint on the causal evolution of the charged components defining the electron (for a modern discussion of the theory see \cite{Spohn2}).
 Another example of theory where maximal acceleration appears is  Caldirola's extended model of charged particles \cite{Caldirola}, where the existence of a maximal speed and minimal elapsed time implies the maximal acceleration.
 However, the notion of  maximal proper acceleration as a fundamental principle of nature was first developed  in
the work of E. Caianiello and co-workers. They provided an heuristic motivation
for such notion in their framework of {\it quantum geometry} \cite{Caianiello}. A different approach was investigated independently by H. Brandt \cite{Brandt1983}.

Caldirola's result on maximal acceleration is a consequence of the extended character of the charged particle in his theory \cite{Caldirola}.
 In contrast, our test particles are  point particles and the maximal acceleration is related with  the assumption of a minimal length $L_{min} $ and maximal speed. The minimal length is assumed to be the  scale of the spacetime region that can produce an effect on the system in the shortest period of time. This notion is not necessarily related with a quantification of spacetime and indeed, the maximal proper acceleration could be relational and depending on the physical system, in contrast with the notion of universal maximal acceleration.

\subsection{On the clock hypothesis and geometry of maximal acceleration}
\
In relativity theory, ideal clocks and rods associated with accelerated particles in relativity are in agreement with the {\it the clock hypothesis}. This assumption is logically independent of {\it the principle of relativity} and {\it the principle of invariance of the speed of light in vacuum respect to any inertial frame}. The hypothesis asserts that for any standard clock and rod, their {\it  behavior depend only upon velocities, and not upon accelerations, or at least, that the influence of acceleration does not counteract that of velocity} (\cite{Einstein1922}, footnote in page 64). However, after the analysis of the related {\it hypothesis of locality} done by B. Mashhoon, it was shown  that  the clock hypothesis is inappropriate when the characteristic acceleration time and length of the phenomena involved are not much larger compared with the intrinsic length and time of the physical system acting as a measurement device \cite{Mashhoon1990}. However, this is exactly the situation for a radiating electron,  where the characteristic acceleration length and time are of the same order than the intrinsic length and time.

A natural way to implement the violation of the clock hypothesis is by using maximal acceleration geometries as the geometry of the spacetime. In a geometry of maximal acceleration the spacetime structure is defined along word lines of accelerated particles.  Hence the standard clocks and roads, depend on the position, velocity and acceleration of the particle \cite{Ricardo2007, Ricardo2015}. We develop below the elements of maximal acceleration theory in classical electrodynamics.

\subsection{Elements of covariant maximal acceleration geometry}
\
If the electromagnetic field is described by a generalized higher order field,
it is natural to describe the spacetime structures by generalized higher order metrics. The justification
for this is two-fold. First, when coupling gravity with
 generalized higher order fields, it is natural to consider the gravitational
 field described by a generalized higher order field. Second, when considering maximal
 acceleration kinematics the corresponding path structure is a generalized metric. In this
  work we apply this second aspect of the theory. We follow mainly the theory of effective metrics of maximal acceleration developed in \cite{Ricardo2015}, to which we refer for the details and proofs of several results stated below.

 Let $M$ be the spacetime manifold. The second jet bundle $J^2_=(M)$ over $M$ is the collection of class of equivalence of smooth curves on $M$ where each class of curves has the same image $x(0)$, the same speed $x'(0)$ and the same acceleration $x''(0)$, namely,
\begin{align*}
J^2_0(M):=\{(x,x',x''),\,x:I\to M \,\textrm{smooth},\, 0\in I\},
 \end{align*}
 where the  coordinates of a given point $u\in\,J^2_0(M)$ are of the general form
 \begin{align*}
 (x,x',x'')=\,\left( x^\mu(t),\frac{dx^\mu(t)}{dt},\frac{d^2 x^\mu(t)}{dt^2}\right) ,\quad \mu=0,1,2,3.
  \end{align*}
  Let $D:\Gamma TM \times\Gamma TM\to \Gamma TM$ be the covariant derivative associated with the Levi-Civita connection of the Lorentzian metric $\eta$. Then we proved in \cite{Ricardo2015} the following
\begin{proposicion}
Let $(M,\eta)$ be a Lorentzian structure,  $x:I\to { M}$  a smooth curve and $T=(x',x'')$ the tangent vector along the lift $x^1:I\to J^1_0(M)$, $t\mapsto(x(t),x'(t))\in \,J^1_0(M)\simeq T_{x(t)}M$ such that $\eta(x',x')\neq 0$ holds.
Then there is a non-degenerate, symmetric form $g$ along $x:I\to M$ such that when acting on the tangent vector $x'(t)$
has the value
\begin{align}
g(\,^2x(t)) (x',x')=\Big(1+ \frac{  \eta(D_{x'}x'(t),
D_{x'}x'(t))}{A^2 _{max}\,\eta(x',x')}\Big)\,\eta(x',x'),
\label{maximalaccelerationmetric0}
\end{align}
where $^2x(t)$ is the second order jet of $x:I \to M$.
\label{teoremasobremaximaacceleration}
\end{proposicion}
The bilinear form $g$  is the {\it metric of maximal acceleration}. The result of its action on two arbitrary vector fields $W,Q$ along $x:I\to M$
is given by
\begin{align*}
g_{^2x(t)}(W,Q)=\,\Big(1+ \frac{  \eta(D_{x'}x'(t),
D_{x'}x'(t))}{A^2 _{max}\,\eta(x',x')}\Big)\eta(W, Q).
\end{align*}
Note that $g$ is not bilinear on the {\it base point} $x'(t)$ but it is bilinear on the vector arguments $W$, $Q$.
 \begin{corolario} Let $x:I \to M$ be a smooth curve such that
 \begin{itemize}
 \item It holds that
 $g(x',x')<0$ and  $\eta(x',x')<0$,
\item The covariant condition
\begin{align}
\eta(D_{x'}\,x',\,D_{x'}\,x')\,\geq 0.
\label{spacelikeaccelerations}
\end{align}
holds good.
 \end{itemize}
Then the bound
 \begin{align}
  0 \leq \eta(D_{x'}x',D_{x'}x')\leq \,A^2_{max}
  \label{boundedconditionforacceleration1}
 \end{align}
 holds good.
 \label{corollary on bound acceleration}
 \end{corolario}
It is now possible to specify in which sense there is a maximal proper acceleration respect to $\eta$.
 For each point $x(t)$ in  the image of a physical world line $x(I)\hookrightarrow M$ and  for any
 instantaneously at rest coordinate system at the point $x(t)\in \, M$, the
 proper acceleration vector field $D_{x'}x'$ along the world line $x:I\to M$ at $x(t)$  is bounded by the maximal proper acceleration $A_{max}$ as indicated by the relation \eqref{boundedconditionforacceleration1}.  A direct consequence is that the relation
\begin{align}
\eta(x'',x'')\leq A^2_{max}
\label{boundproperacceleration}
\end{align}
holds good in any Fermi coordinate system of $D$ along $x:I\to M$.
\subsection{Proper time parameter associated to the metric of maximal acceleration}
We define the proper time associated with $g$ along the world line  $x:I\to M$ with
$\eta(x',x')<0$  by the expression
\begin{align}
\tau[t]=\int^t_{t_0} \,\Big[\Big(1+ \frac{  \eta(D_{x'}x'(s),
D_{x'}x'(s))}{A^2 _{max}\,\eta(x',x')}\Big)(-\eta(x',x'))\Big]^{\frac{1}{2}}\,ds,
\label{propertimeg0}
\end{align}
where $t_0$ is fixed.
Since this expression for the proper parameter $\tau$ is not re-parametrization invariant, we need to fix the parameter
$s$ in a natural way. We choose the parameter $s$ to be the proper time  of $\eta$. Thus the condition
\begin{align*}
\eta(x',x')=-1
\end{align*}
 holds good.
Hence the expression for the proper time of $g$ is
\begin{align}
\tau[t]=\int^t_{t_0} \,\Big[1- \frac{  \eta(D_{x'}x'(s),
D_{x'}x'(s))}{A^2 _{max}}\Big]^{\frac{1}{2}}\,ds.
\label{propertimeg}
\end{align}
As a consequence, it holds that
\begin{align}
\frac{d\tau}{dt}=\big(1-\epsilon\big)^{\frac{1}{2}},
\label{dtaudt}
\end{align}
where the function $\epsilon(t)$ is defined by the expression
\begin{align}
{\epsilon}(t):=\,\frac{\eta(D_{x'}x',D_{x'}x')}{A^2_{max}}.
\label{covariantdefiniciondeepsilon}
\end{align}
 In particular, in a Fermi coordinate system for $\eta$ along $x:I\to M$ the function
$\epsilon(t)$ is given by the expression
\begin{align}
{\epsilon}(t)=\,\frac{ \eta(x''(t),x''(t))}{A^2_{max}}.
\label{definiciondeepsilon}
\end{align}
Note that since there is a bijection $t\mapsto \tau$ given by \eqref{propertimeg0}, it will be more natural to consider the function $\epsilon(\tau)$. Also, by {\it Corollary} \ref{corollary on bound acceleration}, the value  $\epsilon_0$ is finite, even if $I$ is not bounded.
\subsection{Perturbation framework}
The relation between $g$ and $\eta$ along $x:I\to M$ is a conformal factor given by
\begin{align*}
g=\,(1-\epsilon)\eta.
\end{align*}
Since $\epsilon(t)$ is small and we assume that all its derivatives are small too, it determines a bookkeeping parameter ${\epsilon}_0$ by the relation
\begin{align*}
\epsilon_0=\,\max\{\,\epsilon(t),\,t\in I\}.
\end{align*}
Then one can speak of asymptotic expansions on powers of  $\mathcal{O}(\epsilon^l_0)$,
with the basis for the  asymptotic expansions being
\begin{align*}
\{\epsilon^l_0,\, l=-\infty,...,-1,0,1,...,+\infty \}.
\end{align*}

The {\it acceleration square function} is defined by the expression
\begin{align}
 a^2(t):=\,\eta(D_{x'}x',D_{x'}x').
 \label{accelerationsquarefunction}
 \end{align}
 We assume that the dynamics happens in a regime such that
\begin{align}
a^2(t)\ll\, A^2_{max}.
\label{boundedconditionforacceleration}
\end{align}
The curves $X:I\to M$ with $a^2(t)=\, A^2_{max}$ are {\it curves of maximal acceleration}.

 The metric $g$ is the metric that  determines the  proper time measured along the world lines of physical particles. This is the main assumption in the geometry of maximal acceleration, that the physical proper time measured is the one associated to the metric of maximal acceleration $g$. The metric $\eta$ is defined as a formal limit when $A_{max}\to +\infty$, as discussed in \cite{Ricardo2015}.

Let us assume that the
parametrization of the world-line $x:I\to M$ is such that  $g(\dot{x},\dot{x})=-1$ and the monomials in
powers of the derivatives of $\epsilon$ define a complete generator set for asymptotic expansions. Here {\it dot} notation refers to the derivative operation respect to the proper time parameter of $g$. Then one has the relation
 \begin{align*}
 g(\dot{x},\dot{x})=-1\quad \Leftrightarrow\quad \eta(x',x')=-1.
 \end{align*}
This is a particular case of the following relation,
 \begin{align*}
 \eta(x',x')& =\,(1-\epsilon)^{-1}\,g(x',x')=\,(1-\epsilon)^{-1}\,g((1-\epsilon)^{1/2} \dot{x},(1-\epsilon )^{1/2}\dot{x})=\,g(\dot{x},\dot{x}).
 \end{align*}
It is also easy to show that the following kinematical constrains hold,
\begin{align}
& g(\dot{x},\dot{x})=-1,\label{covariantkineticconstrain1}\\
& g(\dot{x},\ddot{x})=\,-\frac{\dot{\epsilon}}{2}+\,\mathcal{O}(\epsilon^2_0) \label{covariantkineticconstrain2}.
\end{align}
The starting point to prove \eqref{covariantkineticconstrain2} is the orthogonality condition in Fermi coordinates of $\eta$,
\begin{align}
\eta(x',x'')=0.
\label{etacommutationrelation}
\end{align}
Together with the relation between the accelerations,
\begin{align*}
\ddot{x}=\,(1-\epsilon )^{-1/2}\,\frac{d}{dt}\left((1-\epsilon)^{-1/2}\right)x'+\,(1-\epsilon)^{-1}\,x''.
\end{align*}
From this expression one can solve for $x''$ and substitute in \eqref{etacommutationrelation},
\begin{align*}
0=\,\eta(x',(1-\epsilon)\ddot{x})-(1-\epsilon)^{1/2}\frac{d}{dt}\left((1-\epsilon)^{-1/2}\right)\eta(x',x').
\end{align*}
Since $\eta(x',x')=-1$, one arrives after some algebra to the expression
\begin{align*}
0& =\,(1-\epsilon)\,\eta(x',\ddot{x})+\,(1-\epsilon)^{1/2}\frac{d}{dt}\left((1-\epsilon)^{-1/2}\right)\\
& =\,(1-\epsilon)^{3/2}\,\eta(\dot{x},\ddot{x})+\,\frac{\dot{\epsilon}}{2}(1-\epsilon)^{-1}\\
& =\,(1-\epsilon)^{3/2}\,\left(g(\dot{x},\ddot{x})+\,\frac{\dot{\epsilon}}{2}(1-\epsilon)^{-5/2}\right),
\end{align*}
from where equation \eqref{covariantkineticconstrain2} follows.
Similar conditions hold for higher derivatives.

\section{A simple derivation of the ALD force equation}

Let us assume that the spacetime is the Minkowski spacetime $(M,\eta)$. In an inertial coordinate  system of $\eta$,
 the ALD equation is the third order differential equation
\begin{equation}
m\,{x}''^{\mu}=\,q\,F^{\mu}\,_{\nu}\,{x}'^{\nu}+\,\frac{2}{3} q^2\,\big({x}'''^{\mu}-
({x}''^{\rho}\,{x}''_{\rho}){x}'^{\mu}\big),\quad {x}''^{\mu}{x}''_{\mu}:={x}''^{\mu}{x}''^{\sigma}\eta_{\mu\sigma},
\label{lorentzdiracequation}
\end{equation}
with $q$ being the charge of the particle, $m$  the experimental
inertial mass and the time parameter being the proper time respect of the metric $\eta$.
The ALD has run-away and pre-accelerated solutions \cite{Dirac,Jackson}, both predictions in contradiction with the first Newton's law of classical dynamics and causality respectively.

We present here an elementary derivation of the ALD equation \eqref{lorentzdiracequation} \cite{Rohrlich}. It illustrates an application of  Cartan's method of moving frames that
we will be adapted later in the contest of generalized higher order fields in combination with maximal
 acceleration geometry. The method does not require the formal introduction of the
 energy-momentum tensor for the electromagnetic field in a explicit way, although it makes use of  the geometry of point particles and the covariant Larmor's law as phenomenological input.

 The formal Lorentz force equation for a {\it bare charged particle} interacting with an electromagnetic field $F_{\mu\nu}$ in a Fermi coordinate system associated to the Lorentzian metric $\eta$ is the equation
\begin{align}
m_b \,{x}''^{\mu}\,=\, q F_{\mu \nu}\, {x}'^{\nu},
\label{non-renormalizedequation}
\end{align}
where $m_b$ is the {\it bare mass}.
Note that both sides in this equation are orthogonal to $x'$ respect to  $\eta$.
In order to generalize the equation \eqref{non-renormalizedequation} to take into account the radiation reaction,
one can add to the right hand side of \eqref{non-renormalizedequation} a vector field $Z$ along the curve $x:I\to { M}$,
\begin{align*}
m_b \,{x}''^{\mu}\,=\, q F_{\mu \nu}\, {x}'^{\nu}+\,Z(\,^3x(t)).
\end{align*}
The orthogonality condition $\eta (Z,x')=0$ implies the following general expression for $Z$,
\begin{align}
& Z^{\mu}(t)=P^{\mu}\,_{ \nu}(t)\big(a_1\,{x}'^{\nu}(t)+a_2\,{x}''^{\nu}(t)+a_3\,{x}'''^{\nu}),\\
&\nonumber  P_{\mu \nu}=\eta_{\mu \nu}+\,{x}'_{\mu}(t){x}'_{\nu}(t),\quad {x}'_{\mu}=\eta_{\mu\nu}\,{x}'^{\nu}.
\end{align}
We can prescribe $a_1=0$. Then using the kinematical relations for $\eta$,
\begin{align*}
x''^\mu\,x'^\nu\,\eta_{\mu\nu}=0,\quad {x}'''^{\rho}\,{x}'^{\sigma}\eta_{\rho\sigma}\,=\,- {x}''^{\rho}\,{x}''^{\sigma}\eta_{\rho\sigma},
\end{align*}
 the relation
\begin{align*}
Z^{\mu}(\tau)=\,a_2 {x}''^{\mu}(t)+\, a_3({x}'''^{\mu}\,-({x}''^{\rho}\,{x}''^{\sigma}\eta_{\rho\sigma})\,{x}'^{\mu})(t)
\end{align*}
is obtained. The term $a_2 \,{x}''$ combines with the left hand side to {\it renormalize} the {\it bare mass},
\begin{align}
(m_b -a_2){x}''^{\mu}=\,m{x}''^{\mu}.
\end{align}
$m$ is the {\it observable mass} or also called the {\it dressed mass}.
This procedure is assumed  valid independently of the values of $m_b$ and $a_2$.

The argument from Rohrlich follows by requiring that the right
 hand side of the equation of motion to be compatible with the relativistic  power radiation formula \cite{Jackson, Rohrlich}. The metric $\eta$ is Minkowski and that Larmor's formula holds good as a consequence of energy-momentum conservation. Therefore, one has
\begin{equation}
\frac{d{p}^{\mu}_{rad}(t)}{dt}=-\, \frac{2}{3}\,q^2 ({x}''^{\rho}\,{x}''^{\sigma}\eta_{\rho\sigma})(t)\,{x}'^{\mu}(t),
\label{larmor}
\end{equation}
where $\frac{d{p}^{\mu}_{rad}(t)}{dt}$ is the rate of radiated $4$-momentum by the particle.
Larmor's law  is satisfied if
 \begin{align*}
  a_3=2/3\,q^2.
 \end{align*}
In order to recover this relation, the minimal piece required in the equation of motion of a
charged particle is $-2/3\, q^2 ({x}''^{\rho}\,{x}''^{\rho}\eta_{\rho\sigma}){x}'^{\mu}$. On the other hand,
the Schott's term $\frac{2}{3}\,q^2{x}'''^\mu$ is a total derivative. It does not contribute to
the averaged power emission of energy-momentum. However, in the above argument the radiation
reaction term and the Schott's term are necessary, due to the kinematical constraints of the metric $\eta$. Then by this procedure the ALD equation is obtained as a description of the dynamics of point charged particles.

\section{A second order differential equation for point charged particles}
\subsection{Larmor's relativistic radiation law in  higher order jet electrodynamics} One of the assumptions that we use in the derivation of the equation of motion of point charged particles is the analogous constraint to  Larmor's relativistic law \eqref{larmor} but formulated in the framework of higher order jets electrodynamics. This assumption however can be motivated in the following way. We expect that in the situation when the radiation reaction is of relevance, the clock hypothesis is not valid. The violation of such hypothesis is mathematically implemented using a metric of maximal acceleration. On the other hand, from a general relativistic point of view, the fact that the metric must be of maximal acceleration implies that in general the fields should also be considered in such category. Thus the electromagnetic field $\bar{F}$ is of the form \eqref{electromagneticfield}, the density current is as in \eqref{generalizedcurrent} and the differential equations that they obey are \eqref{equationforF}, \eqref{equationfor*F} and \eqref{equationforJ}. We assume that $\Upsilon$ is small compared with $F$. Therefore, we are led to consider fields $\bar{F}=\,F+\Upsilon$ such that at leading order are of the form
\begin{align}
\Upsilon\approx \,\epsilon^a \,f(x)+\,\textrm{higher order terms in $\epsilon_0$},
\label{approximationupsilon}
\end{align}
where $a$ is a positive constant. Since the theory is formally the same than in the Maxwell theory,  it is natural to assume that by a similar procedure than in the standard theory the relation
\begin{align}
\frac{d{p}^{\mu}_{rad}}{d\tau}=-\,\left( 1+c_1\,\epsilon^a+\cdot\cdot\cdot\right) \frac{2}{3}\,q^2 (\ddot{x}^{\rho}\,\ddot{x}^{\sigma}g_{\rho\sigma})(\tau)\,\dot{x}^{\mu}(\tau)
\label{generalizedLarmor}
\end{align}
holds good.
$c_1$ is a constant of order $1$.
The first correction factor comes from the energy density for the higher order field.
 We can apply the relation between the proper time parameter $\tau$ of $g$ and the proper time parameter $t$ of the Minkowski metric $\eta$.
 Then \eqref{generalizedLarmor} can be expressed as
\begin{align*}
\frac{d{p}^{\mu}_{rad}}{d\tau}=\,\left( 1+c_1\,\epsilon^a_0+\cdot\cdot\cdot\right)(1-\epsilon)^{-1/2}\,\frac{d{p}^{\mu}_{rad}}{d t}.
\end{align*}
The second correction factor is originated from the factor $(1-\epsilon)$ between $g$ and $\eta$ and its derivatives.
From the definition of $\epsilon$, this can be re-written as
\begin{align*}
\frac{d{p}^{\mu}_{rad}}{d\tau} & =\,\left( 1+c_1\,\epsilon^a_0+\cdot\cdot\cdot\right)(1-\epsilon)^{-1/2}\,\frac{2}{3}\,q^2\,\epsilon\,A^2_{max}\, x'^\mu\\
& =\,\frac{2}{3}\,q^2\,\epsilon\,A^2_{max}\, x'^\mu+\,\textrm{higher order}(\epsilon),
\end{align*}
were the time parameter $t$ is identified with the proper time of the world line of the point particle calculated using the Minkowski metric $\eta$.
Therefore, one has the relation
\begin{align}
\frac{d{p}^{\mu}_{rad}}{d t}=\,\left( 1+c_1\,\epsilon^a_0+\cdot\cdot\cdot\right)\,\left(\frac{2}{3}\,q^2\,\epsilon\,A^2_{max}\, x'^\mu\right).
\label{Lamorapproximateexpressioninhigherorders}
\end{align}
This expression is formally the same than the usual Larmor's formula in Maxwell's theory and shows that at leading order in $\epsilon_0$ in linear higher order field electrodynamics, the standard Larmor's formula must hold.

\subsection{Derivation of the new equation of motion of a point charged particle}
Let us assume that the physical trajectory of a point charged particle is a smooth curve of class $\mathcal{C}^k$
such that $g(\dot{x},\dot{x})=-1$, with $\dot{x}^0<0$ and such that the square of the proper acceleration
$a^2$ is bounded from above.
 We follow closely the analogous argument to Rohrlich's argument as presented in {\it section 4} but in the framework of generalized higher
order fields in a maximal geometry back-ground. If $\bar{F}$ is written in general form \eqref{electromagneticfield}, then
\begin{align*}
\Upsilon_{\mu\nu}(x,\,\dot{x},\,\ddot{x},\,\ddot{x},...)=
\,B_{\mu}\dot{x}_{\nu}\,-B_{\nu}\dot{x}_{\mu}\,+
C_{\mu}\ddot{x}_{\nu}\,-C_{\nu}\ddot{x}_{\mu}\,
+D_{\mu}\dddot{x}_{\nu}\,-D_{\nu}\dddot{x}_{\mu}+...,
\end{align*}
with $\dot{x}_{\mu}=\,g_{\mu\nu}\,\dot{x}^{\nu}$ holds good. This implies an expression for the {\it bare Lorentz force equation} of the form
\begin{align*}
m_b\,\ddot{x}^{\mu} & =\,q\,\bar{F}^{\mu}_\nu\,\dot{x}^\nu=\,q\,F^{\mu}\,_{\nu}\,\dot{x}^\nu+ \big(B^{\mu}\dot{x}_{\nu}-\,\dot{x}^{\mu}B_{\nu}\big)\dot{x}^{\nu}\\
& +\big(C^{\mu}\ddot{x}_{\nu}-\,\ddot{x}^{\mu}C_{\nu}\big)\dot{x}^{\nu} +\big(D^{\mu}\dddot{x}_{\nu}-\,\dddot{x}^{\mu}D_{\nu}\big)\dot{x}^{\nu}\,+...,
\end{align*}
 with $F^{\mu}\,_{\nu}=\,g^{\mu\rho}F_{\rho\sigma}.$
 On the right hand side of the above expression all the contractions that appear in expressions as
 $\big(B^{\mu}\dot{x}_{\nu}-\,\dot{x}^{\mu}B_{\nu}\big)\dot{x}^{\nu}$, etc,  are performed with the metric $g$ instead of the Minkowski metric $\eta$.

The general form of the $k$-jet fields  $B(\tau),C(\tau),D(\tau)$  along the smooth curve $x:\mathbb{R}\to M$ are
\begin{align*}
& B^{\mu}(\tau)=\,\beta_1\,\dot{x}^{\mu}(\tau)\,+\,\beta_2\,\ddot{x}^{\mu}(\tau)\,+
\,\beta_3\,\dddot{x}^{\mu}(\tau)\,+\,\beta_4\,\ddddot{x}^{\mu}(\tau)+\cdot\cdot\cdot,\\
& C^{\mu}(\tau)=\,\gamma_1\,\dot{x}^{\mu}(\tau)\,+\,\gamma_2\,\ddot{x}^{\mu}(\tau)\,+
\,\gamma_3\,\dddot{x}^{\mu}(\tau)\,+\,\gamma_4\,\ddddot{x}^{\mu}(\tau)+\cdot\cdot\cdot,\\
& D^{\mu}(\tau)=\,\delta_1\,\dot{x}^{\mu}(\tau)\,+\,\delta_2\,\ddot{x}^{\mu}(\tau)\,+
\,\delta_3\,\dddot{x}^{\mu}(\tau)\,+\,\delta_4\,\ddddot{x}^{\mu}(\tau)+\cdot\cdot\cdot.
\end{align*}
If one expects that a general relativistic unification point of view is possible, the field $\bar{F}$ and the metric of maximal acceleration should have the same dependence, that is up to second order derivatives of the coordinate world line functions. Therefore, all terms proportional to derivative of higher order than two  must cancel in the above expressions. In particular, the following conditions
\begin{align}
\gamma_k\,=\delta_k\,=0,\, k\geq 0,\quad \beta_k=0, k\geq 3
\end{align}
are shown to be sufficient to our purposes.
With this choice and using the kinetic relations \eqref{covariantkineticconstrain1} and \eqref{covariantkineticconstrain2} for $g$, one obtains the expression
\begin{align}
m_b\,\ddot{x}^{\mu} =\,q\,F^{\mu}\,_{\nu}\dot{x}^\nu-\beta_2 \ddot{x}^{\mu}-\frac{1}{2}\beta_2\,
\dot{\epsilon}\dot{x}^{\mu}.
\label{equationbeforerenormalization}
\end{align}
Under the assumption  that the equation is compatible with the relativistic Larmor power radiation formula \eqref{larmor} and for $\dot{\epsilon}({s})\neq 0$,
 one obtains the relations
\begin{align}
&\beta_2=\, \frac{4}{3}q^2\,a^2(\tau)\,\frac{1}{\dot{\epsilon}}.
\end{align}
Note that in the coefficient $\beta_2$ it appears $a^2(\tau)=\eta(x'',x'')$, instead of $g(\ddot{x},\ddot{x})$. However, it is direct that at leading order these two terms are equivalent.
At leading order in $\epsilon_0$, we obtain the following differential constraint:
\begin{align}
m_b(\tau)\,\ddot{x}^{\mu} =\,q\,F^\mu\,_\nu\dot{x}^\nu\,-\frac{2}{3}\,q^2\,a^2(\tau)\,\dot{x}^{\mu}-
\,\frac{2}{3}\,q^2\,a^2(\tau)\,\frac{1}{\dot{\epsilon}}\,\ddot{x}^{\mu}.
\label{prerenormalization}
\end{align}
There is a re-normalization of the bare mass,
\begin{align}
 m_b(\tau)+\,\frac{2}{3}\,q^2\,a^2(\tau)\,\frac{1}{\dot{\epsilon}}=\,m,\quad \dot{\epsilon} \neq 0.
\label{renormalizationofmass}
\end{align}
In this expression, the term $m_b$ contains already the divergent term
from the Coulomb field, that is the infinite electrostatic mass has been
already renormalized (for instance the term $\ddot{x}^\mu C_\nu\dot{x}^\nu$ is a term that can renormalize the Coulomb self-energy).
As a consequence of the above reasoning, for $\dot{\epsilon}\neq 0$ the following implicit differential equation holds,
\begin{align*}
m\,\ddot{x}^{\mu} =\,q\,F^{\mu}\,_{\nu}\,\dot{x}^{\nu}-
\,\frac{2}{3}\,{q^2}\,\eta_{\rho\sigma}\,{x}''^{\rho}{x}''^{\sigma}\,\dot{x}^{\mu},\quad F^{\mu}\,_{\nu}=\,g^{\mu\rho}F_{\rho\nu}.
\end{align*}
As in the discussion of the Larmor's formula in higher order electrodynamics, one can show that this expression is at leading order term equivalent to
\begin{align}
m\,\ddot{x}^{\mu} =\,q\,F^{\mu}\,_{\nu}\,\dot{x}^{\nu}-
\,\frac{2}{3}\,{q^2}\,\eta_{\rho\sigma}\,\ddot{x}^{\rho}\ddot{x}^{\sigma}\,\dot{x}^{\mu},\quad F^{\mu}\,_{\nu}=\,g^{\mu\rho}F_{\rho\nu}.
\label{equationofmotion}
\end{align}
Note that we have written $a(\tau)$ instead of $a(t)$. This is a valid substitution, since introduce a higher error in $\epsilon$ only. Equation \eqref{equationofmotion} is our new equation of motion.

Note that the derivation of the implicit second order differential equation \eqref{equationofmotion} breaks down for $\dot{\epsilon}=0$, a case that we will consider separately. Also, it implies that the generalized electromagnetic field $\bar{F}$ depends  of the second order jets, consistent with the definition of the metric of maximal acceleration.

In the equation \eqref{equationofmotion}, time derivatives are taken respect to the proper parameter $\tau$ associated to the metric of maximal acceleration \eqref{maximalaccelerationmetric0}. This opens the natural question of the observability and falsifiability of the equation \eqref{equationofmotion}. However, note that in our theory of metrics with maximal acceleration, the parameter $\tau$ is the proper time measured by a co-moving coordinate system associated to the charged point particle. A natural strategy to see that equation \eqref{equationofmotion} is falsifiable is the following. First, one writes down the equation in the co-moving coordinate system associated to the charged particle. This allows to select the proper time $\tau$ as the parameter time of the equation. Then the equation in such coordinate system is solved. Third, one  transforms the solutions to usual inertial coordinate systems associated to laboratory frames. This type of procedure avoids to change the parameter $\tau$ to $t$ in the expression \eqref{equationofmotion}, avoiding the use of the parameter $t$ in the theory except for the theoretical construction of the differential equation. Moreover, since the equation is Lorentz covariant,  in order to check the expression in an inertial frame associated to a laboratory system, it is only required to know the instantaneous position and speed of the point particle respect to the inertial frame. With such information, the local coordinate representation in the inertial coordinate system of the world line of the point particle can be constructed using instantaneous local Poincar\'e transformations.

The general covariant form of equation \eqref{equationofmotion} in any coordinate system is
\begin{align}
m\,D_{\dot{x}}\,\dot{x}=\,q\,(\iota_{\dot{x}}F)^*(x(\tau))\,-\frac{2}{3}\,q^2\eta(D_{\dot{x}}\,\dot{x},D_{\dot{x}}\,\dot{x})\,\dot{x},
\label{covariantequationofmotion}
\end{align}
with $(\iota_{\dot{x}}F)^*$ the dual of the $1$-form $\iota_{\dot{x}}F\in \,\Gamma\Lambda^1 M$.
Note that since we are considering
  metrics of maximal acceleration avoids the orthogonality problem of taking away the Schott's term in the ALD equation.

\subsection{Properties of the equation \eqref{equationofmotion}}

Let us define
\begin{align*}
(F_L)^\mu=\,q((\iota_{\dot{x}}F)^*)^\mu =\,qF^\mu\,_\nu(x)\,\dot{x}^\nu,\quad \mu=0,1,2,3.
\end{align*}
\begin{proposicion}
If the Lorentz force is zero, then for any curve solution of equation \eqref{equationofmotion} the magnitude of the acceleration is zero,
\begin{align}
F^2_L=0\,\quad \Rightarrow\quad  \,a^2=0.
\label{conditiononceroaccelerationandceroexternalforce}
\end{align}
\label{propiedadesequationofmotion}
\end{proposicion}
\begin{proof}
Let us multiply equation \eqref{equationofmotion} by itself and contract each side with the metric $g$,
\begin{align*}
m^2\,(g(\ddot{x},\ddot{x}))^2  =\,g(F_L,F_L)+(2/3 \,q^2)^2 (g(\ddot{x},\ddot{x}))^2\,g(\dot{x},\dot{x})-\,(2/3,q^2)\,g(\dot{x},\dot{x})\,g(\ddot{x},F_L).
\end{align*}
Since $g(\dot{x},\dot{x})=-1$ and if $F_L=0$, we have that
\begin{align*}
m^2\,(g(\ddot{x},\ddot{x}))^2=\,-(2/3 \,q^2)^2 (g(\ddot{x},\ddot{x}))^2.
\end{align*}
The left hand side is positive or zero. On the other hand, the right hand side is negative or zero. Hence both sides must be zero and then $g(\ddot{x},\ddot{x})=0$. Since the vector field $\ddot{x}$ is spacelike or zero respect to $\eta$, it must be also spacelike or zero respect to $g$. Then it must be zero.
\end{proof}
{\it Run away} solutions have the following peculiar behavior. Even if the
external forces have a compact domain in the spacetime, the charged particle follows
accelerating indefinitely. The original ALD equation \eqref{lorentzdiracequation} has run away solutions for a dense set of initial conditions, except if Dirac's asymptotic condition is imposed \cite{Dirac,Spohn2}. Although the asymptotic condition has a natural interpretation within  the framework of singular perturbation theory and it is related with Landau-Lifshitz theory, it  still leaves open the question of the existence of pre-accelerated solutions, since we are imposing a kinematical constrain on the future of the motion of the particle.
In contrast, since the condition \eqref{conditiononceroaccelerationandceroexternalforce} holds for equation \eqref{equationofmotion}, we show that our model is free of run away solutions.

The relation \eqref{equationofmotion} is an {\it implicit differential equation}.
 If the radiation reaction term $\theta=\frac{2}{3}q^2 a^2 x^\mu$ is small compared with
 the Lorentz force term $F_L$, it can be treated  as a perturbation. Then at leading order in the product $\theta\cdot\,\epsilon$, one can consider that the Lorentz force  equation is a good approximation,
 \begin{align*}
 \ddot{x}^\mu=\,\frac{q}{m}\,F^\mu\,_\rho\dot{x}^\rho+\,\mathcal{O}(\theta\,\cdot\epsilon)
 \end{align*}
  and substitute this expression in the equation \eqref{equationofmotion}, keeping the leading order zero in the product $\mathcal{O}(\theta\cdot\epsilon)$. This leads to the approximate equation
   \begin{align}
m\ddot{x}^\mu=\,q F^\mu\,_\nu\,\dot{x}^\nu-\,\big(\frac{2\,q^4}{3m^2}\big) \,\dot{x}^\mu \big(F_{\rho\lambda}\dot{x}^\lambda F^\rho\,_\sigma\dot{x}^\sigma\big).
\label{approximationofequationofmotion}
\end{align}
   The error in such approximation is of higher order  $\theta \cdot\,\epsilon$, each factors assumed to be small. Therefore, for formal purposes equation \eqref{equationofmotion} can be substituted by \eqref{approximationofequationofmotion}.

   Note that of the equation of motion, the second derivative $\ddot{x}$ is isolated. Therefore,
for this approximated equation \eqref{approximationofequationofmotion}
one can apply existence and uniqueness theorem of ODE theory \cite{Hartman} to prove the following consequence,
\begin{proposicion}
Let $z_1$ and $z_2$ be two solutions of the equation \eqref{equationofmotion} with the same initial
conditions. Then they differ locally by a small function of the expression $\theta\cdot\epsilon$.
\end{proposicion}
\begin{proof} Let us consider the solutions $z_1$ and $z_2$ of equation \eqref{equationofmotion} with the same initial conditions. They can be approximated by one solution $\tilde{z}$ of equation of the equation \eqref{approximationofequationofmotion}, since the unique initial conditions. Furthermore, the difference in the solutions $\tilde{z}^\mu-z^\mu_1$ and $\tilde{z}^\mu-z^\mu_2$ is controlled by $\theta\cdot \epsilon$. Thus for small values of the difference $\theta\cdot \epsilon$, there is a small difference of the corresponding solutions of the equations \eqref{equationofmotion} and \eqref{approximationofequationofmotion}. By uniqueness of solutions of ODE \cite{Hartman}, given a fixed initial conditions the solution to \eqref{approximationofequationofmotion} exits and is unique (in a finite interval of time $\tau$). In such interval the two possibly different  solutions of the original equation \eqref{equationofmotion} differ by a small function of the expression $\theta\cdot \epsilon$. \end{proof}
 \subsection{Value of the maximal acceleration in classical electrodynamics}
If we multiply the left side of the differential equation \eqref{approximationofequationofmotion} by itself and the right side of the equation by itself and  contract each side by itself with the metric $\eta$, then at the same order of approximation one obtains the relation
\begin{align*}
m^2\,a^2=\,F^2_L\left(1-\,\left(\frac{2}{3}\,\left(\frac{q}{m}\right)^2\right)^2\,F^2_L\right).
\end{align*}
Since the left hand side is never negative, the same must be true for the right hand side. Thus there is a maximal value for the strength of the electromagnetic field, namely,
\begin{align}
F^2_L\leq \,\left(\frac{2}{3}\,\left(\frac{q}{m}\right)^2\right)^{-2}.
\end{align}
This implies a bound for the acceleration that a point charged particle can reach by means of an external electromagnetic field,
\begin{align}
a\leq \frac{3}{2}\,\frac{m}{q^2}=A_{max}.
\label{valueofthemaximalacceleration}
\end{align}
This value of the maximal acceleration coincides with the same maximal acceleration in Caldirola's theory  \cite{Caldirola}.

The specific value of the maximal acceleration depends on the particle with charge $q$ and mass $m$. For an electron, it is of order $A_{max}=10^{32} m/s^2$, but this bound can be considerably lower for non-neutral plasmas must happen in accelerator physics. In particular, in a rough model of a plasma as a point particle with charge $q=N e$ and $m=N m_e$, it is of the form
\begin{align}
a\leq \frac{3}{2}\,\frac{m_e}{e^2}\,\frac{1}{N}\sim\,\frac{1}{N}\,10^{32}\,ms^{-2},
\end{align}
where $e$ is the charge of the electron, $m_e$ is the mass of the electron and $N$ is the number of particles in the plasma. It is currently possible to have bunches with $N\sim 10^9$ or bigger bunches of particles in particle accelerators, in which case $A_{max}$ could be of order $10^{21}$-$10^{22}$ $m/s^2$. This order of acceleration could be also reached in laser-plasma accelerators. In particular, in \cite{Litos} it was reported acceleration of order $10^{20}$ $m/s^2$ for a charge satem of $74$ pico-Coulomb (corresponds to $N\sim 10^8$). If the bunch is to be described by a unique particle, then the maximal acceleration of the system must be of order $10^{24}$ $m/s^2$. These  estimates suggest the possibility to measure effects of maximal acceleration in classical electrodynamics in laser-plasma acceleration.

\subsection{Covariant uniform acceleration}

The case of world-lines with $\dot{\epsilon}=0$ requires special care. Let
us consider the following definition of {\it covariant uniform acceleration},
\begin{definicion}
A covariant uniform acceleration curve is a map $x:I\to M$ such that along it the following constrain holds,
\begin{align}
\dot{\epsilon}=\frac{d}{d\tau}\frac{\eta(D_{{x}'}{x}',D_{{x}'}{x}')}{A^2_{max}}=0.
\label{equatioofdefinitionofuniformacceleration}
\end{align}
\label{definitionofuniformacceleration}
\end{definicion}

Equation \eqref{equationofmotion} fails to describe covariant uniform motion. The reason
is that, in order to apply  the mass renormalization procedure  \eqref{renormalizationofmass}
and the perturbation scheme, the condition  $\dot{\epsilon}\neq 0$ was applied. Therefore,
one needs to consider separately the case $\dot{\epsilon}=0$ and re-derive the equation of motion following an analogous procedure. Let us consider the following consequence of $\dot{\epsilon}=0$ in the relation \eqref{equationbeforerenormalization},
\begin{align*}
m_b \,\ddot{x}^{\mu}=-\,qF^\mu\,_\nu\dot{x}^\nu+\tilde{\beta}_2\ddot{x}^\mu.
\end{align*}
Thus for covariant uniform motion, the following equation follows,
\begin{align}
m\ddot{x}^\mu=\,q\,F^\mu\,_\nu\dot{x}^\nu,\quad m=m_b(\tau)-\,\tilde{\beta}_2(\tau).
\label{covariantequationofmotionform2}
\end{align}
This is the Lorentz force equation for the external electromagnetic field $F_{\mu\nu}$. The conflict between the \eqref{covariantequationofmotionform2} and the equation \eqref{equationofmotion} is resolved by adopting the point of view that it cannot be ideal hyperbolic motion caused by electromagnetic means only. The existence of back-reaction prevents the point particle of being in uniform acceleration, if no other forces besides electromagnetic forces are applied.

\subsection{Absence of pre-acceleration of Dirac's type for the equation \eqref{equationofmotion}}
Let us consider the example of a pulsed electric field
\cite{Dirac}. In Fermi coordinates of $\eta$, for an external electric pulse $\vec{E}=(\kappa\,\delta(\tau),0,0)$,
the equation $(\ref{equationofmotion})$ in the non-relativistic limit reduces to
\begin{align}
a\ddot{x}^0=\kappa\,\delta(\tau),\quad a=\frac{3m}{2q^2},
\label{solutiondiracpulse}
\end{align}
the maximal acceleration associated. The solution of this equation is the {\it Heaviside's function},
\begin{align*}
a\dot{x} = &  \kappa,\,\tau\geq 0,\\
&  0,\,\tau< 0,
\end{align*}
that does not exhibit pre-acceleration behavior. In order to prove uniqueness on the space of
smooth functions, let us consider two solutions to  the equation \eqref{solutiondiracpulse}.
Clearly, the two solutions must differ by an affine function of $\tau$ and the requirement of having the same
initial conditions, implies that the affine function must be trivial. This shows that
in the non-relativistic limit the equation $(\ref{equationofmotion})$ does not have pre-accelerated solutions of Dirac's type, but since the theory is Lorentz covariant, the equation $(\ref{equationofmotion})$ does not have such kind of solutions in
 any coordinate system.
It is open the question if $(\ref{equationofmotion})$ is free of any other type of pre-accelerated solutions.

\section{Discussion}

Combining the notion of generalized higher order electromagnetic field with maximal acceleration geometry
we have obtained the implicit differential equation \eqref{equationofmotion} as  description of the dynamics
of a point charged particle.
 Equation \eqref{equationofmotion} and its general covariant version equation \eqref{covariantequationofmotion} is free of run-away solutions and pre-accelerate solutions of Dirac's type.
 The assumption of maximal acceleration is necessary in order to keep under control the value of the acceleration.
 It also provides a book-keeping parameter $\epsilon_0$ useful to construct our perturbative theory. The hypothesis of
 generalized higher order fields is fundamental too, since these fields provide the degrees of freedom that we need
to eliminate the Schott's term in the ALD equation. Moreover, the   physical interpretation of such fields
  is quite natural, indicating that electromagnetic fields should not be defined independently of the way in which they are measured or detected, even at the classical level description.

The generalized fields are assumed to be well defined on the trajectory of the test particle.
This is contradictory with the fact that electromagnetic fields are divergent at the localization
in spacetime of the point charged particle world-line, due to the Coulomb singularity. In order
to treat this problem we have adopted the method of mass renormalization, with a time variable bare mass $m_b(\tau)$. The method allows us to consider well defined fields
over the world-line of charged particle probes, without the singularity of Coulomb type, which is renormalized
to give the observable constant mass $m$. Therefore, the domain of definition of our model is a local domain of spacetime manifold $M$ where the particle moves, but a restricted domain on the second jet $J^2(M)$, according with the perturbative restrictions of the model.

Quite remarkably, the model implies a  concrete value of the maximal acceleration. This value depends on the characteristics of the particle or system. Because such dependence on the specific characteristics of the particle, it is possible to test the theory due to its implications for the dynamics of non-neutral plasmas. For the same reason, the theory presented has potential applications in modelling the dynamics of relativistic non-neutral plasma systems and in laser-plasma acceleration.

\appendix

\section{Jet bundles and generalized tensors}

In this appendix we collect several notions and definitions that we need
 to introduce generalized higher order fields. An extended version of the theory
is developed in \cite{Ricardo2012} and the interested reader is referred there for more details. The differential geometry of jets is found thoroughly developed in \cite{KolarMichorSlovak}.
Given a smooth curve $x:I\to { M}$, the collection
$(p=x(0),\,\frac{dx}{d\sigma},\,\frac{d^2x}{d\sigma^2},..., \frac{d^kx}{d\sigma^k})$ determines a point (jet) in the space of
jets $J^k_0(p)$ over the point $p\in \,M$,
\begin{align*}
J^k_0(p):=\left\{(x(0),\,\frac{dx}{d \sigma}\big|_{0},\,...,\,\frac{d^k x}{d\sigma^k}\big|_{0}),
\,\,\forall\,\mathcal{C}^{k}\textrm{-curve}\,\,\, x:I\to M,\, x(0)=p\in M,\,0\in I \right\}.
\end{align*}
The set of higher derivatives $(p=x(0),\,\frac{dx}{d\sigma},\,\frac{d^2x}{d\sigma^2},..., \frac{d^kx}{d\sigma^k})$ determines a $k$-jet at the point $p=x(0)$.
 The jet bundle $J^k_0(M)$ over $M$ is the disjoint union
\begin{align*}
J^k_0(M):=\bigsqcup_{x\in M}\,J^k_0(x).
\end{align*}
The canonical projection map is
\begin{align*}
^k\pi  :J^k_0(M)\to M,\quad
 (x(0),\,\frac{dx}{d\sigma}\big|_0,\,\frac{d^2x}{d\sigma^2}\big|_0,..., \frac{d^k x}{d\sigma^k}\big|_0)\mapsto x(0).
\end{align*}
The simplest example where jets appear is when we use Taylor's expansions of smooth functions. This kind of approximation is an example of the type of approximation.

An important result of jet theory is {\it Peetre's theorem}. Recall that the support $supp$ of a {\it section} $S:M\to \mathcal{E}$ of a vector bundle is the closure of the sets $\{x\in \,M;\,S(x)\neq 0\}$. An operator $D$ between the bundles $\pi_1:\mathcal{E}_1\to \,M$ and $\pi_2:\mathcal{E}_2\to\,M$ $D:\mathcal{C}^\infty(\mathcal{E}_1)\to\mathcal{C}^\infty(\mathcal{E}_2)$ is said to be support non-increasing if $supp(D S)\subset supp(S)$ for every section $S\in \,\Gamma \mathcal{S}$.

\begin{teorema}[Peetre, 1960] Consider two vector bundles $\pi_1:\mathcal{E}_1\to \,M$ and $\pi_2:\mathcal{E}_2\to\,M$ and a non-increasing operator $D:\mathcal{C}^\infty(\mathcal{E}_1)\to\mathcal{C}^\infty(\mathcal{E}_2)$. For every compact set  $K\subset\,supp D$, there is an integer $r(K)$ such that if the jets $j^r(S_1)=j^r(S_2),$ then $D S_1=\,D S_2$.
\end{teorema}

The following definition provides the fundamental notion of generalized higher order field,
\begin{definicion}
A generalized tensor $T$ of type $(p,q)$ with values on $\mathcal{F}(J^k_0(M))$ is a smooth section of the bundle of $\mathcal{F}(M)$-linear homomorphisms
\begin{align*}
T^{(p,q)}(M,\mathcal{F}(J^k_0(M))) :=\,Hom(T^*M\times...^p...\times T^*M\times TM\times ...^q...\times TM,\,\mathcal{F}(J^k_0(M))).
\end{align*}
A $p$-form $\omega$ with values on $\mathcal{F}(J^k_0(M))$ is a smooth section of the bundle of $\mathcal{F}(M)$-linear completely alternate homomorphisms
\begin{align*}
\Lambda^p(M,\mathcal{F}(J^k_0(M))) :=\,Alt(TM\times ...^p...\times TM,\,\mathcal{F}(J^k_0(M))).
\end{align*}
The space of $0$-forms is $\Gamma\,\Lambda^0(M,\mathcal{F}(J^k_0(M))):=\mathcal{F}(J^k_0(M))$.
\label{definiciontensoerFJMvaluados}
\end{definicion}
\begin{ejemplo}
A generalized electromagnetic field is a section of $\Lambda^2(M,\mathcal{F}(J^3_0(M)))$.
\end{ejemplo}

There is also defined a Cartan's  calculus is naturally defined for generalized forms. It is naturally constructed in close analogy to  the
standard Cartan's calculus of smooth differential forms. The usual formulation of the inner derivation,
exterior product and exterior derivative can be introduced in a coordinate free manner and are independent of the connection used (for the construction and general properties of Cartan calculus of generalized forms, the reader can see \cite{Ricardo2012}). For instance, there is a coordinate free definition of the exterior derivative operator $d_4$. The realization in local coordinates of the operator $d_4$ is straightforward: if $\phi$ is a generalized $k$-form, then
\begin{align*}
d_4\phi & =d_4\big(\phi_i(x,\,\dot{x},\,\ddot{x},..., x^{(k)}) d_4x^i\big)\\
& = d\big(\phi_i(x,\,\dot{x},\,\ddot{x},..., x^{(k)})\big)\wedge\, d_4x^i\\
& =\partial_j\,\phi_i(x,\,\dot{x},\,\ddot{x},..., x^{(k)})\,d_4x^j\,\wedge d_4x^i.
\end{align*}
This operator is nil-potent, $(d_4)^2=0$. Other analogous properties of the usual exterior derivative $d$ also holds in the case of the operator $d_4$.

\subsection*{Acknowledgement} We acknowledge  financial support from FAPESP, process 2010/11934-6, the Riemann
Center for Geometry and Physics from the Leibniz University Hanover, Germany and  PNPD-Capes n. 2265/2011, Brazil, at different stages of this research.
\small{
}

\end{document}